\documentclass[]{article}
\usepackage{graphicx}
\usepackage{placeins}
\usepackage[utf8]{inputenc}
\usepackage{bookmark}
\usepackage{amsmath}
\usepackage{multirow}
\usepackage{amssymb}
\usepackage{pdfpages}
\usepackage{xcolor}
\usepackage{authblk}
\usepackage{listings}

\providecommand{\keywords}[1]
{
  \small	
  \textbf{\textit{Keywords---}} #1
}

\begin{document}
\title{A model of communication-enabled traffic interactions}

\author[1]{O. Siebinga\footnote{Corresponding author: o.siebinga@tudelft.nl}}
\author[1]{A. Zgonnikov}
\author[1]{D.A. Abbink}

\affil[1]{Delft University of Technology - Department of Cognitive Robotics \\ Mekelweg 2, Delft, the Netherlands}

\date{April 2023}

\maketitle
\keywords{Driving Interactions, Driver Modelling, Traffic Communication}

\begin{abstract}
A major challenge for autonomous vehicles is handling interactive scenarios, such as highway merging, with human-driven vehicles. A better understanding of human interactive behaviour could help address this challenge. Such understanding could be obtained through modelling human behaviour. However, existing modelling approaches predominantly neglect communication between drivers and assume that some drivers in the interaction only respond to others, but do not actively influence them. Here we argue that addressing these two limitations is crucial for accurate modelling of interactions. We propose a new computational framework addressing these limitations. Similar to game-theoretic approaches, we model the interaction in an integral way rather than modelling an isolated driver who only responds to their environment. Contrary to game theory, our framework explicitly incorporates communication and bounded rationality. We demonstrate the model in a simplified merging scenario, illustrating that it generates plausible interactive behaviour (e.g., aggressive and conservative merging). Furthermore, human-like gap-keeping behaviour emerged in a car-following scenario directly from risk perception without the explicit implementation of time or distance gaps in the model's decision-making. These results suggest that our framework is a promising approach to interaction modelling that can support the development of interaction-aware autonomous vehicles.
\end{abstract}

\section{Introduction}
Autonomous vehicles (AVs) hold the potential to help address major societal challenges related to mobility and sustainability. However, one of the major open problems in autonomous vehicle development is safely and acceptably dealing with driving scenarios that require \textit{two-way interaction} with human road users. In these interactions, such as in highway merging or intersection negotiation, both vehicles reciprocally influence and respond to the actions of each other. It entails quick and sometimes iterative negotiations, based on communication (see e.g.,~\cite{Potzy2019,Ulbrich2015,Lee2021}) that can either be implicit (vehicle motions) or explicit (e.g., honking, signalling). The continuous dynamics of a two-way interaction govern safety, priority (who goes first, who gives way), and acceptance (by passengers and other road users). For example, drivers can be misunderstood or cause annoyance by being too conservative or aggressive (interfering with, or ignoring others' communication). Therefore, fundamental knowledge about continuous human two-way interactions is necessary to develop and evaluate safe and acceptable AV behaviour for these scenarios. However, this fundamental knowledge about the dynamics of interactions is currently lacking. We advocate using a modelling approach for human two-way traffic interactions to develop the fundamental understanding that in the future can help design better AV behaviour.

Modelling is a common way of gaining an understanding of human driving behaviour. But it has so far mostly been done with a focus on single-driver behaviour, either in single-vehicle (e.g.,~\cite{Kolekar2020, Salvucci2004}), or multi-vehicle scenarios such as car following~\cite{Treiber2008, Kesting2010}, lane changing~\cite{Rahman2013,Salvucci2002}, and gap acceptance~\cite{Zgonnikov2020, Tian2022}. Most multi-vehicle approaches assume that the modelled driver responds to other traffic participants, but that they don't respond in turn. For example, car-following models assume that the following driver responds to the leading vehicle, but this leading vehicle does not change its behaviour based on the follower's actions. We call this the \textit{one-way interaction} assumption. This assumption disentangles the behaviours of the multiple drivers and thereby enables the researchers to better understand and model the behaviour of the driver of interest. The scope of these models is thus deliberately restricted to a single driver. This one-way interaction assumption is justified for car-following models and the likes, but not for interactive driving scenarios like merging or intersection negotiations, which are inherently reciprocal. Simply joining two one-way interaction models to describe an interaction will neglect the drivers' beliefs about the other's future actions and their expected influence on it. Furthermore, it also neglects the presence and effects of communication between the drivers. Therefore, we argue that the scope of an interaction model should include all participants to begin with.

The current mainstream approach to modelling complete traffic interactions (as opposed to individual drivers) is using game theory. Game theory was developed as a framework to describe two-way interactions between players in abstract games. It has been used extensively to model traffic interactions. The first model of human merging behaviour based on game theory was proposed in 1999 by Kita~\cite{Kita1999}. In 2007, Liu et al. improved the game theoretical approach by removing the assumption of constant velocity~\cite{Liu2007}. After that, many works followed (e.g.~\cite{Meng2016, Tian2019, Zhang2018, Coskun2019}). However, applying game theory to model dynamics between two drivers is not trivial, because game theory makes three strong assumptions about these players. 

First, the assumption that all players rationally maximize some utility function. Empirical evidence has shown that even in simple economic games~\cite{Camerer2003}, but also in driving behaviour~\cite{Schmidt-Daffy2014} and traffic interactions~\cite{Kalantari2023}, this assumption does not hold for human players. Second, game theory does not allow communication between the players, an aspect known to be important in interactive driving scenarios~\cite{Lee2021}. Third, the majority of game-theory-based interaction models use a set of discrete actions for the drivers.  Although this is useful to describe the higher-level tactical~\cite{Michon1985} decisions of drivers accurately (for example the decision to yield or merge), it does not describe the lower-level operational~\cite{Michon1985} dynamics of the interaction (e.g. changes in velocity or trajectory). Therefore, these approaches are not sufficiently detailed for developing safe and acceptable AV behaviour. Combined, these three limitations motivate the need for an alternative approach to modelling two-way traffic interactions that allows for communication, bounded rationality, and continuous dynamic actions.

To address this gap, here we propose a framework for Communication-Enabled-Interaction (CEI) modelling. It can be used to create model implementations, of which we provide one example in a case study\footnote{The software implementation of the presented model and its simulation environment are available online at~\cite{Siebinga2022}. The data discussed in the results section can be found at~\cite{Siebinga2022e}.}. The modelling framework relaxes the common assumptions that drivers are rational agents and have full information about the strategies of other drivers. It is based on the notion that all drivers have a plan they want to execute and a belief about what other drivers are going to do. Combined, this plan and belief result in a perceived risk for every driver. The drivers are assumed to act to keep this risk below their individual threshold. The key insight of the framework is that the beliefs about others are updated based on communication between the agents. In a simulation case study, we show that an implementation of a CEI-model produces plausible behaviour of two interacting drivers in a simplified merging scenario. Besides that, human-like gap-keeping behaviour emerges directly from the notion of risk perception. These results show that the proposed modelling framework provides a promising new approach for modelling human-human driver interactions.

\section{Communication-Enabled Interaction (CEI) \\ Modelling}
\begin{figure}[h]
    \centering
    \includegraphics[width=\textwidth]{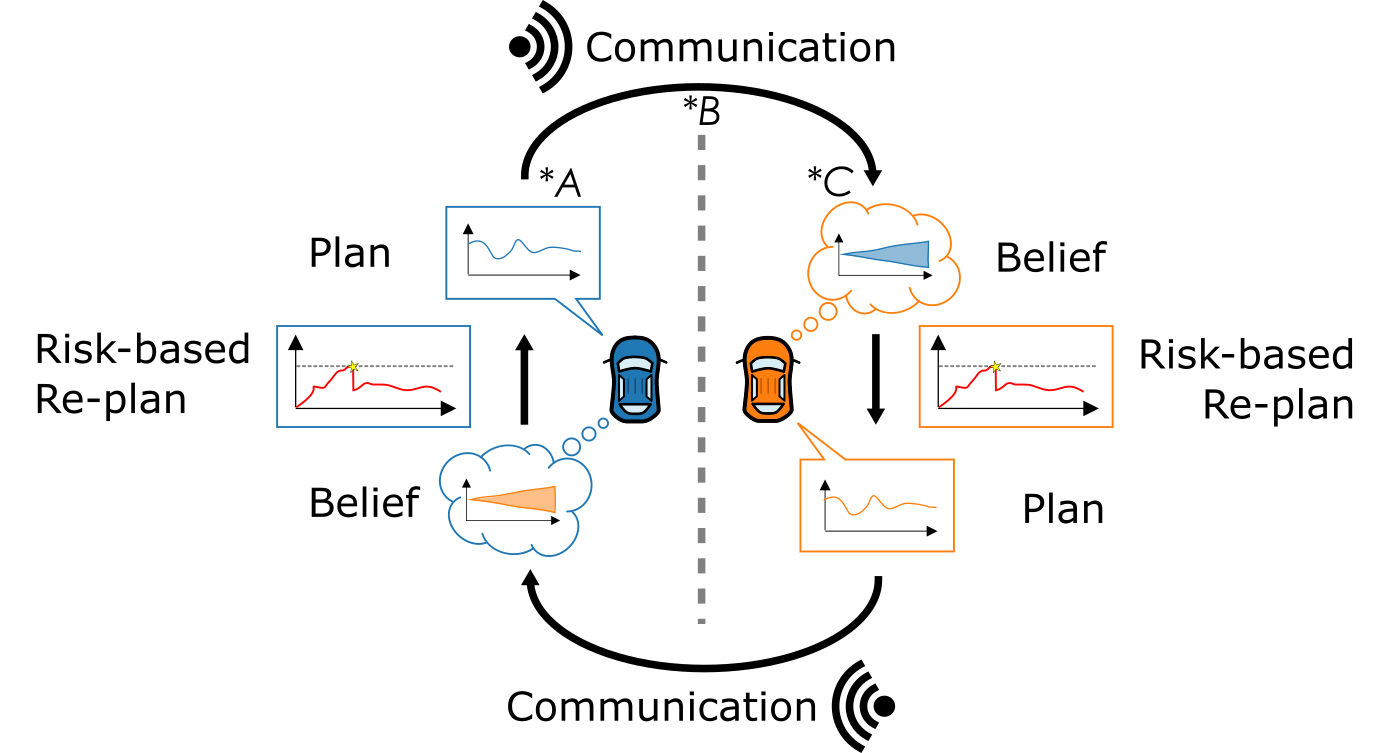}
    \caption{An overview of the proposed Communication-Enabled-Interaction (CEI) modelling framework. This framework is designed to capture the two-way interaction between two drivers, rather than the one-way interaction behaviour of one driver with respect to another. Each driver has a \textit{plan} for their own behaviour. Plan updates are triggered based on a \textit{risk threshold} and a \textit{risk estimate} arising from a \textit{belief} of how the other driver will move over time. Each driver \textit{communicates} their plan (intention) either implicitly (e.g., through vehicle motion), or explicitly (e.g., through light signals) to the other driver. This communication links one driver's plans to the belief of the other and can be divided into three components denoted \textit{*A}, \textit{*B}, and \textit{*C}. \textit{*A} represents the mapping of a driver's plan to its communication, \textit{*B} represents the means of communication, and \textit{*C} denotes the belief update of the other driver based on the received communication.}
    \label{fig:model}
\end{figure}

We propose a framework to model human-human traffic interactions between two drivers. This framework puts the modelling scope around the complete interaction rather than a single driver, and explicitly includes communication between the drivers. Each driver is described by four components: a notion of risk, a deterministic plan (for their own behaviour), a means of communication, and a probabilistic belief about the future actions of the other driver (Figure~\ref{fig:model}). The general framework we present here only defines loose requirements for these components. When implementing the model for a specific scenario or use case, these components can be designed based on existing literature (e.g., from the fields of human behaviour modelling, traffic communication, intent inference, or vehicle path planning). The advantage of this is that one can leverage knowledge from the literature to improve the model, without having to fully redesign it. In this section, we will discuss the four components and our reasoning behind them. The assumptions and requirements that need to be taken into account when implementing a model based on this framework will also be discussed per component. In Section~\ref{sec:methods}, we will illustrate how each component can be implemented in an example implementation for a simplified merging scenario.

\subsection{Framework components}

\subsubsection*{Risk-Based Re-plan}
Recent research has shown that risk plays an important role in human driving behaviour~\cite{Kolekar2020, Kolekar2020b, Jensen2022}. In our framework, we combine this notion of risk-based decision making in driving with Simon's ideas of \textit{bounded rationality}~\cite{Simon1955} and \textit{satisficing}~\cite{Simon1956}. \textit{Bounded rationality} implies that humans are not capable of fully optimizing their behaviour all the time. \textit{Satisficing} (a portmanteau of \textit{satisfy} and \textit{suffice}) is an example of bounded rationality in which humans are assumed to not continuously search for an optimal solution. Instead, they are \textit{satisfied} with a "good enough" solution that \textit{suffices}. We reason that the only solutions that suffice and satisfy in a driving interaction are the ones that are \textit{subjectively safe enough}. To formalize these ideas, and combine them in a framework, we hypothesize that drivers act to keep their perceived risk below their risk threshold. 

Using such a threshold incorporates Simon's ideas in two ways. First, it defines what solutions are subjectively safe enough. Second, it limits (or bounds) the cognitive capacities (or effort) required from the driver because it allows the driver to only rethink their plan when the situation changed and the current plan does not suffice or satisfy anymore. This is what we call a \textit{risk-based re-plan} (Figure~\ref{fig:model}). By incorporating these ideas, we step away from the fundamental assumption of game theory that humans are rational utility maximizers and move towards a formulation that allows for team effort and mutual goals. 

In summary, our framework assumes every driver to evaluate the risk of their current deterministic plan, given their probabilistic belief about what other drivers are planning to do. Risk perception can be based on a number of factors, such as high velocity, high acceleration, or the probability of a collision. This evaluation happens continuously, but drivers will only perform a re-plan if the perceived risk exceeds their threshold. This should result in drivers with a low risk threshold adapting their plan in an early stage of the interaction to reduce the estimated risk. At the same time, drivers with a high risk threshold will instead continue their current plan and take advantage of the fact that the risk of the situation is lowered by the other driver. Intuitively this can be explained as the driver with the higher risk threshold being more aggressive.

\subsubsection*{Plan}
The second component in our framework is the \textit{plan}. We assume that drivers have a deterministic plan about the actions they will take in the immediate future. In the framework, this plan takes the form of a deterministic set of waypoints over a limited time horizon. This time horizon should be long enough to include (part of) the interaction. 

The construction of this plan (i.e., the planning algorithm) should only consider features that are not related to risk and safety (e.g., desired velocity or comfort), as the perceived risk is constantly evaluated separately to determine if the current plan still suffices and satisfies. This evaluation is done taking into account both the plan and the belief. When re-planning, the risk threshold should be used as a constraint in the planning algorithm. As long as such a constraint can be imposed, the plan can be constructed using any suitable path-planning algorithm.

\subsubsection*{Communication}
One of the key concepts of the framework is that drivers actively communicate their plan to other drivers. This assumption is based on field studies on human-human traffic interaction that confirm that traffic participants actively communicate their plan both explicitly and implicitly to others (e.g.~\cite{Lee2021}). Experiments on other (non-driving) tasks that require team effort have shown that humans use their movement actions to coordinate with their team member~\cite{Sacheli2013} (which is a form of implicit communication). The assumption of communication can also be effectively used to model human behaviour in those tasks~\cite{Pezzulo2013}. Finally, in simulation, communication can be beneficial for controlling co-bots that navigate among humans~\cite{Dadvar2021}, resulting in fewer dead-lock situations. In summary, previous research suggests that humans communicate in traffic and that the assumption of communication can be used both for the effective modelling of human teamwork behaviour and the effective control of robots.

In the CEI modelling framework, communication links the plan of one driver to the belief of the other driver. In practice, this means that three aspects of communication need to be designed when implementing a CEI-model. First, one needs to determine the mode of communication; What signals are used to communicate? These signals can be explicit (e.g., turn indicators) or implicit (e.g., velocity, heading angle, or acceleration). Second, a mapping from a plan to its communication is required. This can be as simple as just executing the plan, but one could come up with more elaborate mappings based on traffic communication studies such as slowing down, purely to communicate that the other driver can go first (for an example of modelling such exaggerated trajectories in a bottle grasping task, see~\cite{Pezzulo2013}). Finally, a mapping from communication to belief is needed, this mapping specifies how a probabilistic belief is updated based on the received communication.

\subsubsection*{Belief}
Both drivers are assumed to have probabilistic beliefs about what the other driver will do in the near future. This belief consists of a number of points over a time horizon. Each of these belief points is represented by a probability distribution over positions for the other driver for that specific time in the future (Figure~\ref{fig:model}). This assumption is based on the intuition that human drivers have a general but uncertain idea about what other drivers are planning to do, a concept that has been successfully applied in other modelling frameworks such as belief-desire-intention programming (based on~\cite{Bratman1987}) and (Bayesian) theory of mind~\cite{Baker2011} as well. 

When implementing the belief part of the CEI-model, the only requirement is that the chosen probability distribution can be updated using new information (coming from the observed communication). In practice, this means that most parametric probability distributions are suitable because they can be updated with methods such as Bayesian updates.  

\section{Case Study: an Example of an Implementation}
\label{sec:methods}
To demonstrate the feasibility of the proposed model framework and to investigate the effects of design choices (parameters) on model behaviour, we have implemented a CEI-model for a simplified merging scenario. In this case study, we show that even with simple components the model framework can produce plausible, human-like interactive behaviour. At the same time, it is not the purpose of this case study to quantitatively assess the model's consistency with human behaviour. Such an assessment using fine-grained data on the interactive behaviour of two drivers requires a detailed investigation and is therefore left for future work.

\subsection{Simplified merging scenario}

\begin{figure}[h]
    \centering
    \includegraphics[width=0.99\textwidth]{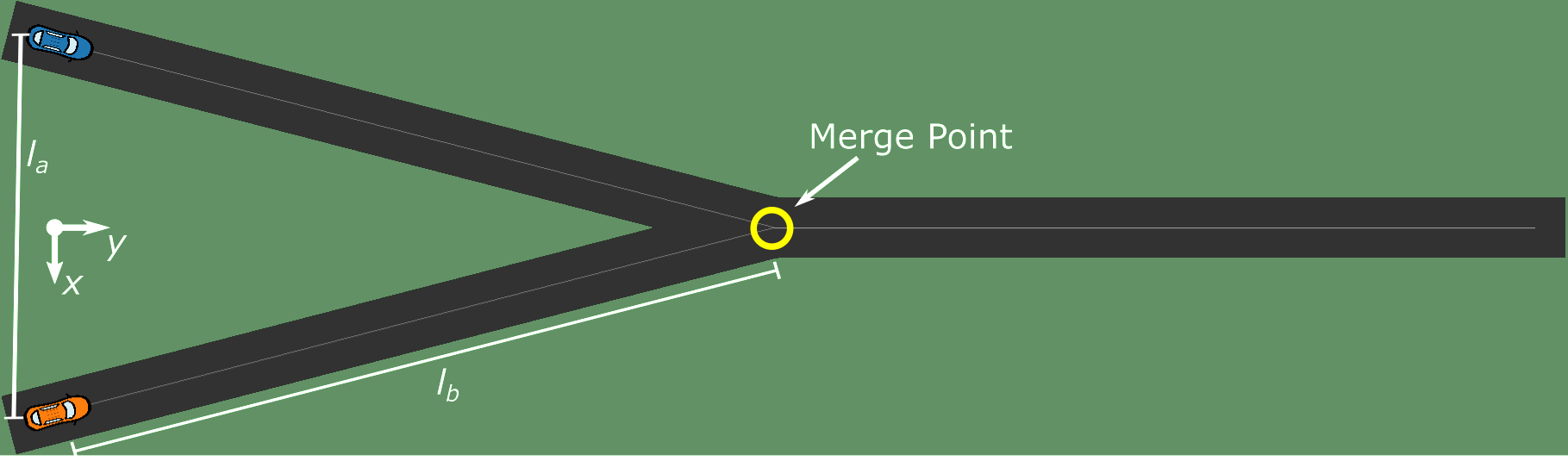}
    \caption{A top-down view of the simplified merging scenario as used in the case study, rotated 90 degrees clockwise. Vehicles follow pre-defined paths (road centres) that merge at a pre-defined merge point. Vehicles have a two-dimensional body ($4.5~m$ x $1.8~m$) and their headings change instantly at the merge point. The model controls the accelerations of the vehicles directly. The dimensions of the track are defined by two parameters. Distance $l_a$ ($25~m$) denotes the distance between the start points of the vehicles. Distance $l_b$ ($50~m$) is the distance to travel from the start point until the merge point, and from the merge point until the end of the track.}
    \label{fig:scenario}
\end{figure}

For this case study, we used a simplified symmetric merging scenario (Figure~\ref{fig:scenario}). In this scenario, two vehicles approach a merge point on a predefined track. The model can directly control the acceleration of the vehicles, but there is no steering involved. The vehicles have a rectangular bounding box for collision detection. The heading of the vehicles is pre-defined and always corresponds to the heading of the road. At the merge point, the heading of the vehicles changes instantly. 

The vehicles in the simplified scenarios are subject to a negative acceleration due to resistance and drag. The net acceleration ($a^{net}$) is the applied input ($a^{in}$) minus the negative acceleration $a^r$ (a function of the vehicle's velocity $v$):

\begin{align}\label{eq:resistance}
    a^{net}(v) &= a^{in} - a^r(v) \text{, where} \\
    a^r(v) &= \alpha v^2 + \beta.
\end{align}

Parameters $\alpha$ and $\beta$ define the magnitude of the drag and constant resistance ($\alpha = 0.0005$ and $\beta = 0.1$). Besides the resistance, the vehicles have a maximum acceleration $a^{max} = 2.5~\frac{m}{s^2}$, which is the same for positive and negative accelerations. The velocity of the vehicles is restricted to non-negative values. The simulation updates all dynamics at a rate of $20~Hz$. 

\subsection{Plan}
The planning part of the model consists of a path planning algorithm that minimizes the following cost function:

\begin{equation}
\label{eq:cost}
    c = \sum^N (v_n - v^d)^2 + (a^{in}_n)^2.
\end{equation}

Where $n$ denotes the time-step and $v$ the vehicle's velocity. This cost function includes terms for minimizing the squared input $a^{in}$ and for travelling at a desired velocity $v^d$. The path is planned at the same frequency as the simulation ($20~Hz$) and is subject to a time horizon of $4~s$ ($N=\frac{4}{0.05}=80$). 

\begin{figure}
    \centering
    \includegraphics[width=\textwidth]{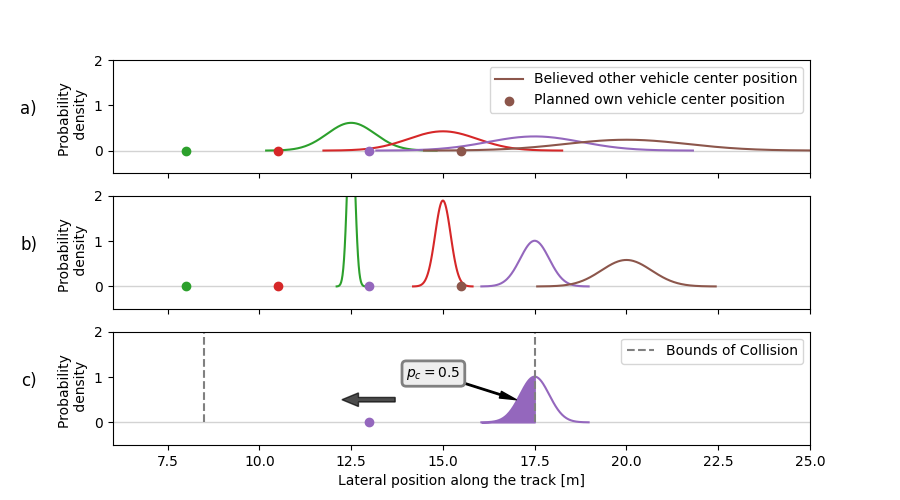}
    \caption{An example to illustrate the plan and the belief of the model. a) shows four (of 80) deterministic plan points along the one-dimensional track. These are the planned centre positions of the own vehicle at four points in time. The distributions represent the believed centre position of the other vehicle at the same four (of 16) points in time, where colours denote the points in time. b) shows these plan and belief points after a single belief update. This update increased the certainty of the belief about the other vehicle's position. The belief is updated at every time step. c) shows the risk evaluation for one of the points. To evaluate the risk, the probability of a collision ($p_c$) is evaluated by calculating the probability that the other vehicle will be within the bounds of collision for the given planned position. This risk evaluation is done at every time step for all belief points. If the maximum perceived risk value exceeds the upper risk threshold, a re-plan is triggered. This re-plan uses the perceived risk as a constraint for the optimization. To lower the risk, the planned position could be moved in the direction of the black arrow.}
    \label{fig:plan_and_belief_example}
\end{figure}

A visual example of the plan, belief, and risk perception is shown in Figure~\ref{fig:plan_and_belief_example}. When initially planning the path, the cost function of Equation~\ref{eq:cost} is minimized, so an optimal path is found with respect to comfort and speed (Figure~\ref{fig:plan_and_belief_example}-A). If, at the next time step, the current plan still satisfies (i.e., the risk threshold is not exceeded), the current plan is continued. We assume that maintaining velocity at the final time step is the practical equivalent of maintaining the current plan.

When the risk threshold is exceeded, the cost function is minimized again to find a new plan (Figure~\ref{fig:plan_and_belief_example}-C). This time the minimization is subject to a risk constraint. Based on the ideas of satisficing, we hypothesise that humans do not spend unlimited effort to find an optimal plan, but instead search for a new solution that satisfies and suffices. We hypothesize that re-planning is easiest (i.e., requires the least cognitive effort) if the new plan is close to the previous plan (i.e., uses the same strategy). Therefore, the re-planning optimization is executed with the old plan as the initial condition. When using a gradient descent algorithm, this will result in a solution that is close to the previous plan while the risk constraint is met. For example, if the current plan is to decelerate and pass behind the other driver, the most likely outcome of the re-planning will be to decelerate even more and increase the gap. This will lower the perceived risk while using the current strategy. 

If the optimization with the current plan as the initial condition does not succeed, three other initial conditions are considered: full braking at all time steps, no acceleration input at all time steps, and full acceleration at all time steps. The candidate plan with the lowest cost is used as the initial condition for a second re-plan. This can result in a change of strategy, but only if the current strategy is not feasible anymore. For example, when the driver was decelerating but decelerating even more will not reduce the risk enough, it will investigate if acceleration will reduce the risk and change its strategy if needed.

\subsection{Belief}
The belief is kept as a sequence of probability distributions over positions for the other vehicle, each at a specific point in time (Figure~\ref{fig:plan_and_belief_example}-A). This sequence of belief points uses the same time horizon as the planning part of the model ($4~s$) but contains fewer points for simplicity. Belief points are kept at a $4~Hz$ frequency (this number was based on an initial evaluation of the model), resulting in a sequence of $4 \cdot 4=16$ points. Each belief point is represented by a Gaussian distribution. 

The Gaussian distributions are initialized by combining the initial velocity and position of the other vehicle with the maximum bounds of acceleration. To initialize a belief point, the mean of the Gaussian is set to the position that corresponds to the other driver maintaining its current velocity. To calculate the standard deviation, an upper and lower position bound ($ub$ and $lb$) are used. These are calculated by predicting the position of the other vehicle if it would apply the maximum and minimum possible acceleration continuously. The standard deviation is then calculated as the difference between the bounds and the mean divided by 3 ($\sigma = \frac{ub - \mu}{3}$). The factor $\frac{1}{3}$ is based on the fact that $99.73\%$ of the area under a normal distribution corresponds to $\mu +/- 3\sigma$. Once the simulation time is equal to the timestamp corresponding to the first belief point, this point is removed from the sequence and a new point is initialized.

\subsection{Communication}
Human communication during driving is a complex topic on which a lot of research has been done. Thus, there is much potential for including complex communication models based on empirical evidence in a CEI-model. However, for this initial investigation of the modelling framework, we used a simple implicit communication model that does not include any explicit communication signals (e.g., turn indicators). We only use velocity and position as communication signals. These two values are assumed to be constantly observed by the other driver without any errors or noise. 

When sending communication, the drivers do not use a mapping from their current plan to the actions they take. Instead, they just take the next action from their plan. When receiving communication, drivers use a constant velocity model combined with bounds of comfortable acceleration to update their belief. All belief points are updated every time step using Bayesian updating. 

\subsubsection{Updating the Belief}
For Bayesian updating, the previous belief point serves as the prior distribution, and the resulting posterior is adopted as the updated belief point (Figure~\ref{fig:plan_and_belief_example}-B). The likelihood is constructed using the constant-velocity model. We assume the likelihood to be a Gaussian distribution where the standard deviation is constant and known. This means the likelihood and prior form a conjugate pair, meaning that the posterior will also be a Gaussian distribution of which the $\mu$ and $\sigma^2$ have a closed-form solution. The likelihood function for the belief point at time $t$ is defined as follows: 

\begin{equation}
    \mathcal{N}\left(\mu=\frac{p}{t}, \sigma^2=\left(\frac{a_c t}{6}\right)^2\right)
\end{equation}

In this equation, $p$ denotes a position sampled from the prior (the previous belief point), $t$ denotes the time corresponding to the belief point, and $a_c$ is the maximum comfortable acceleration ($a_c=1.0~\frac{m}{s}$). The same value is used for positive and negative accelerations, thus the distribution is symmetrical. The likelihood function describes the probability of observing a velocity $v$ (now) given a sampled predicted position $p$ (at time $t$) from the prior belief. The mean $\mu$ corresponds to constant velocity, and $\sigma$ is determined based on the assumption that $99.73\%$ of the distribution falls within the bounds of comfortable acceleration. 

With this likelihood function, the posterior has a closed form solution. We denote the prior as $\mathcal{N}(\mu_0, \sigma^2_0)$ and the posterior as $\mathcal{N}(\mu_1, \sigma^2_1)$. When updating with a single data point $v$, the solution for the posterior becomes\footnote{For a complete derivation of this closed-form solution, see the supplementary material.}:

\begin{align}
    \mu_1 &= \frac{\mu_0 \sigma^2 + v \sigma^2  \frac{1}{t}}{\sigma^2 + \sigma^2_0  \frac{1}{t^2}} \\ 
    \sigma^2_1 &= \frac{\sigma^2 \sigma^2_0}{\sigma^2 + \sigma^2_0  \frac{1}{t^2}} 
\end{align}

\subsection{Risk}
\label{sec:risk}

The risk perceived by the drivers is assumed to be proportional to the probability of a collision. Other aspects (i.e., high velocity and high acceleration) are assumed not to contribute to the perceived risk for simplicity. To estimate the probability of a collision, we define the concept of \textit{bounds of collision} (Figure~\ref{fig:plan_and_belief_example}-C). These are the extreme positions of the other vehicle that would result in a collision, given the position of the own vehicle. These bounds are calculated for every point in the driver's plan. For example, if we know the driver will be at position $x$ at time $t$, we can use the vehicles' dimensions to calculate that a collision will occur if and only if the other vehicle is at a position between $x+c_1$ and $x-c_2$ at the same time; these are the bounds of collision. The believed probability that the other vehicle will be within these bounds at that time can be calculated using the belief about the other vehicle's position. This probability is then equal to the probability of a collision at that time.  

The perceived risk for a complete plan is determined by taking the maximum risk over all belief points. A re-plan is triggered if the perceived risk exceeds an upper threshold $\rho_u$. Only using the upper threshold, however, poses a potential problem when the merging conflict is resolved because after that there will be no triggers to re-plan anymore. This might cause vehicles to stall or drive very slowly for no reason. We avoid this by extending the risk module with a lower risk threshold $\rho_l$ and a saturation time $\tau$. If the perceived risk is lower than $\rho_l$ and the last update was longer than $\tau$ ago, a re-plan is also triggered. When a re-plan optimization is performed, the perceived risk is constrained to be lower than the average of the two thresholds. For the implementation of this constraint, the instant heading change at the merge point in the track posed a problem. Therefore, a linear approximation of the bounds of collision is used. 

\subsection{Investigated Scenarios}
In total, every driver in the model has four parameters that determine their behaviour: a desired velocity $v_d$, an upper risk threshold $\rho_u$, a lower risk threshold $\rho_l$, and a saturation time $\tau$. Besides these parameters, the initial velocity and position ($v_0$ and $x_0$) of the drivers can also be adjusted. Both drivers always start from the beginning of the track. In the case study, we investigate the effect of these parameters and the effect of differences in the initial condition in four scenarios (Table~\ref{tab:scenario_parameters}). 

\begin{table}[!h]
\centering
\caption{Parameters of the investigated scenarios. Underlined values denote deviations from the default values. $\rho_l$ and $\rho_u$ denote the lower and upper risk thresholds. $v_0$ and $v_d$ are the initial and desired velocity respectively. $x_0$ denotes the initial position of the vehicle along the track.}
\label{tab:scenario_parameters}
\begin{tabular}{c|llllll|}
\cline{2-7}
\multicolumn{1}{l|}{} &  \multicolumn{1}{l|}{Side} &  \multicolumn{1}{l|}{$\rho_l$} &  \multicolumn{1}{l|}{$\rho_u$} &  \multicolumn{1}{l|}{$v_0$} &  \multicolumn{1}{l|}{$v_d$} &  $x_0$ \\ \hline

\multicolumn{1}{|l|}{Units} &  \multicolumn{1}{l|}{-} &  \multicolumn{1}{l|}{\textbf{-}} &  \multicolumn{1}{l|}{\textbf{-}} &
  \multicolumn{1}{l|}{$\frac{m}{s}$} &  \multicolumn{1}{l|}{$\frac{m}{s}$} &  $m$ \\ \hline
\multicolumn{1}{|c|}{\multirow{2}{*}{\begin{tabular}[c]{@{}c@{}}Condition A:\\ No expected collision\end{tabular}}} &
  left &  $0.2$ &  $0.5$ &  $10.0$ &  $10.0$ &  $0.0$ \\
\multicolumn{1}{|c|}{} &  right &  $0.2$ &  $0.5$ &  $\underline{\mathbf{9.0}}$ &  $\underline{\mathbf{9.0}}$ &  $0.0$ \\ \hline
\multicolumn{1}{|c|}{\multirow{2}{*}{\begin{tabular}[c]{@{}c@{}}Condition B:\\ On a collision course\end{tabular}}} &
  left &  $0.2$ &  $0.5$ &  $10.0$ &  $10.0$ &  $0.0$ \\
\multicolumn{1}{|c|}{} &  right &  $0.2$ &  $0.5$ &  $\underline{\mathbf{9.0}}$ &  $\underline{\mathbf{9.0}}$ &  $\underline{\mathbf{1.2}}$ \\ \hline
\multicolumn{1}{|c|}{\multirow{2}{*}{\begin{tabular}[c]{@{}c@{}}Condition C:\\ High and low thresholds\end{tabular}}} &
  left &  $0.2$ &  $\underline{\mathbf{0.4}}$ &  $10.0$ &  $10.0$ &  $0.0$ \\
\multicolumn{1}{|c|}{} &  right &  $\underline{\mathbf{0.3}}$ &  $\underline{\mathbf{0.6}}$ &  $10.0$ &  $10.0$ &  $0.0$ \\ \hline
\multicolumn{1}{|c|}{\multirow{2}{*}{\begin{tabular}[c]{@{}c@{}}Condition D:\\ Threshold sensitivity\end{tabular}}} &
  left &  $\underline{\mathbf{0.3}}$ &  $\underline{\mathbf{0.4}}$ &  $10.0$ &  $10.0$ &  $0.0$ \\
  \multicolumn{1}{|c|}{} &  right &  $\underline{\mathbf{0.3}}$ &  $\underline{\mathbf{0.6}}$ &  $10.0$ &  $10.0$ &  $0.0$ \\ \hline
\end{tabular}
\end{table}

The first two scenarios (A \& B) manipulate the initial and desired velocity of the right driver while keeping the parameters of the left driver fixed; the drivers here have the same risk thresholds. In scenario A, the drivers are not expected to be on a collision course if they would stick to their desired velocity, but in scenario B, they are. 

Scenarios C \& D focus on the risk thresholds. Scenario C investigates the effect of a difference in risk thresholds between drivers. Scenario D investigates the sensitivity of model behaviour to variations of these thresholds in one of the drivers. The saturation time $\tau$ only affects the behaviour after the conflict is resolved, therefore it is kept constant at $2.0~s$ for all scenarios. 

\section{Results}
\subsection{Scenario A: No expected collision}

\begin{figure}[ht!]
    \centering
    \includegraphics[width=0.9\textwidth]{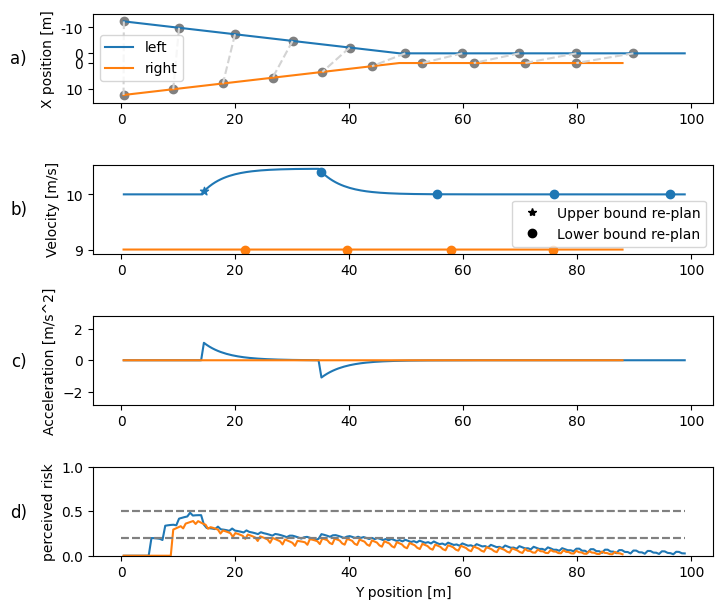}
    \caption{Model behaviour in scenario A (no expected collision). Line colours correspond to the vehicle colours in Figure~\ref{fig:scenario}. a) Positions of the left and right vehicles over time. The x positions of the vehicles are plotted with an offset to prevent the lines from overlapping after the merge point. The grey dots and dashed lines indicate vehicle positions at equal time stamps with an interval of $1.0~s$. b) Velocities of the vehicles over time. The stars indicate the moment when the simulated driver performed a re-plan because the upper risk threshold was exceeded, and a circle denotes a re-plan because the risk fell below the lower threshold. These re-plans are only triggered if the last re-plan was longer than $\tau$ ago. c) Accelerations of the vehicles over time. d) Perceived risk of both simulated drivers. In case of a re-plan, the perceived risk after the re-plan is shown. The dashed horizontal lines in the lowest plots indicate the risk thresholds of the drivers. In this scenario, the drivers increased the small projected gap, even though they were initially not on a collision course. The simulated drivers behaved in a way to increase the initially narrow safety margin.}
    \label{fig:scenarioA}
\end{figure}

Scenario A serves as a baseline scenario. Here, both drivers have an initial velocity that is equal to their desired velocity, but that differs from the velocity of the other driver (Table~\ref{tab:scenario_parameters}). If they would keep their initial (desired) velocity up until the merge point, no collision would occur. The left driver would pass the merge point first with a small distance gap of $0.2~m$. Therefore we would expect a rational optimizing model (that does not explicitly include human-like gap-keeping) to maintain the desired velocity all the way. A behaviour expected from human drivers, on the other hand, is to increase this small safety margin. In an empirical study~\cite{Daamen2010}, it was found that human drivers in the Netherlands merged on three different highway locations with mean headways of $12.6$, $13.4$, \& $36.1~m$ for velocities below $60~km/h=16.7~m/s$, and standard deviations of respectively $10.3$, $12.8$ \& $18.2$ (the headway is defined as the gap plus the leading vehicle length).

In the modelled outcome of scenario A (Figure~\ref{fig:scenarioA}), the left driver reached the merge point first. They accelerated slightly to increase the safety margin at the merge point, after that, they returned to their preferred velocity. The headway when the second vehicle reached the merge point was $6.4~m$. This corresponds to the expected human behaviour, and can not be modelled with utility-maximization unless utility is explicitly awarded for keeping a gap. The right driver did not take any action in this scenario. The reason for that is highlighted in the risk perception plot. The left driver's risk increases earlier because it expects to reach the merge point earlier. This increase causes the left driver to take action to lower the risk, while the right driver can continue their plan without exceeding their risk threshold. The right driver's perceived risk also decreases as soon as the left driver takes action; they perceive that the conflict was resolved by the left driver.

\FloatBarrier
\subsection{Scenario B: On a collision course}

In scenario B, the drivers have the same desired and initial velocities as in scenario A. However, the right vehicle starts with a $1.2~m$ head-start. Therefore, the projected positions of the two vehicles at the merge point overlap by $1.0~m$. Thus, if neither driver deviates from their desired velocity, this scenario will result in a collision. We would therefore expect that this scenario requires more severe action to be resolved than scenario A, but we do expect the model to avoid a collision. 

\begin{figure}[h!]
    \centering
    \includegraphics[width=0.9\textwidth]{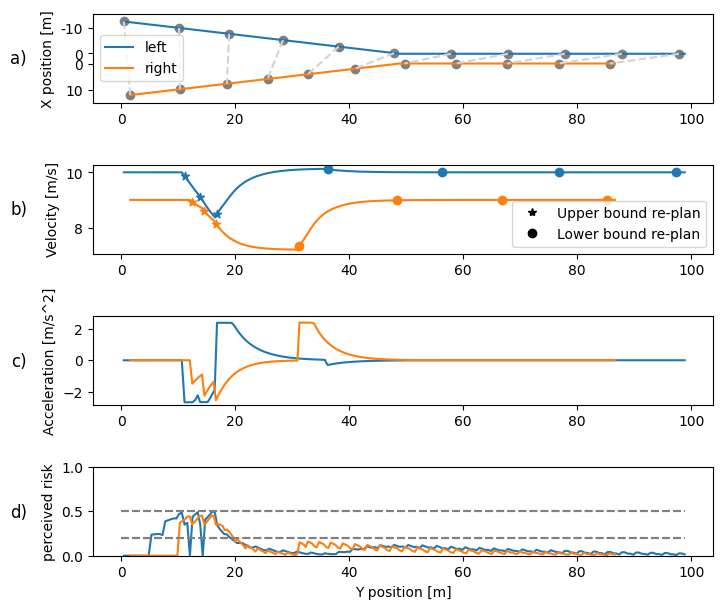}
    \caption{Model behaviour in scenario B. The simulated drivers prevent a collision by slowing down. Initially, they both slow down, but after approximately one second, the left (initially faster) driver speeds up and reaches the merge point first. For details of the notation, see the caption of Figure~\ref{fig:scenarioA}.}
    \label{fig:scenarioB}
\end{figure}

The modelled outcome of scenario B (Figure~\ref{fig:scenarioB}) shows that this scenario indeed requires more effort from both drivers to resolve the conflict compared to scenario A. Both drivers start braking until the left driver decides they can only reduce the risk of a collision by accelerating. This can be explained by the fact that the left driver has a slightly higher velocity at this point compared to the right driver. The right driver sticks to their plan and keeps decelerating until the risk drops below the lower threshold and the saturation time has passed, only then they accelerate again. This behaviour results in a safety margin between the vehicles that is not explicitly included in the reward function. Because the left driver is the first to accelerate, they reach the merge point first. This explainable interactive behaviour combined with the collision-free outcome can be regarded as a plausible human-like interaction. 

\subsection{Summary Scenarios A and B}
In scenario A, the driver with the higher preferred velocity that approached the merge point first also passed the merge point first. But the distance gap between the vehicle was enlarged by the drivers. This corresponds to what we expected from human drivers. If the drivers approach the merge point with an expected collision (scenario B), however, the drivers take more drastic action but still manage to resolve the conflict by interacting with each other. 

\FloatBarrier
\subsection{Scenario C: High and low thresholds}
Scenario C represents a case where the simulated drivers of both vehicles have the same initial conditions and desired velocities, but different risk thresholds. Compared to the previous scenarios, the right driver has higher risk thresholds while the left driver has lower thresholds. The left driver, having lower thresholds, is expected to act early in the interaction to reduce their perceived risk. In terms of human behaviour, this would correspond to risk-averse, conservative driving. The right driver (high thresholds meaning higher tolerance to risk) is expected to react to a potential conflict at a later point and therefore to keep their velocity at the desired level longer. We expect that the right driver reaches the merge point first, and deviates less from their desired velocity compared to the left driver.

\begin{figure}[h!]
    \centering
    \includegraphics[width=0.9\textwidth]{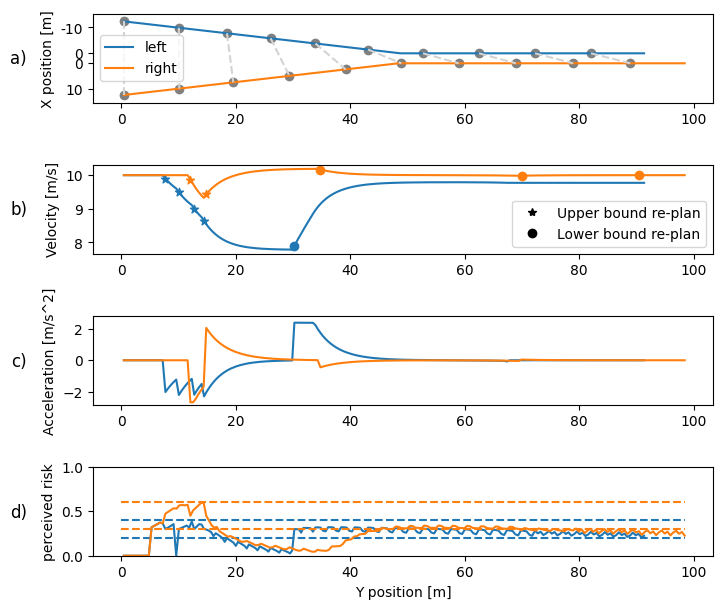}
    \caption{Model behaviour in scenario C. The right driver maintains their initial velocity longer. After briefly decelerating, they accelerate and reach the merge point first. For details of the notation, see the caption of Figure~\ref{fig:scenarioA}.}
    \label{fig:scenarioC}
\end{figure}

The modelled outcome of scenario C (Figure~\ref{fig:scenarioC}) is as expected: the left driver reached their upper threshold first and started to decelerate to reduce the perceived risk. In terms of human driving, this can be seen as more conservative behaviour. The right driver reacts later because their risk threshold is exceeded at a later moment. They briefly decelerate, but quickly start to accelerate to reduce the risk since the left driver already decelerated. This results in the right driver reaching the merge point first and deviating less from their desired velocity than the left driver. This corresponds to the intuition that lower sensitivity to risk (i.e. higher risk thresholds) could be associated with more aggressive behaviour. 

\subsection{Scenario D: Threshold sensitivity}
Scenario D investigates the sensitivity of the modelled drivers' behaviour to variations in the lower risk threshold. This scenario is the same as scenario C, with the only exception that the left driver has a slightly higher value for $\rho_l$ (lower risk threshold). We, therefore, expect a very similar outcome in scenarios C and D. The only expected difference is that the left driver in scenario D re-plans more frequently because the risk for the new plan is constrained to the average of the two risk thresholds. With a smaller difference between $\rho_l$ and $\rho_u$, the absolute risk decrease at the re-plan points is smaller. This should cause the perceived risk to reach the upper threshold quicker and thus result in more frequent re-plan events.

\begin{figure}[h!]
    \centering
    \includegraphics[width=0.9\textwidth]{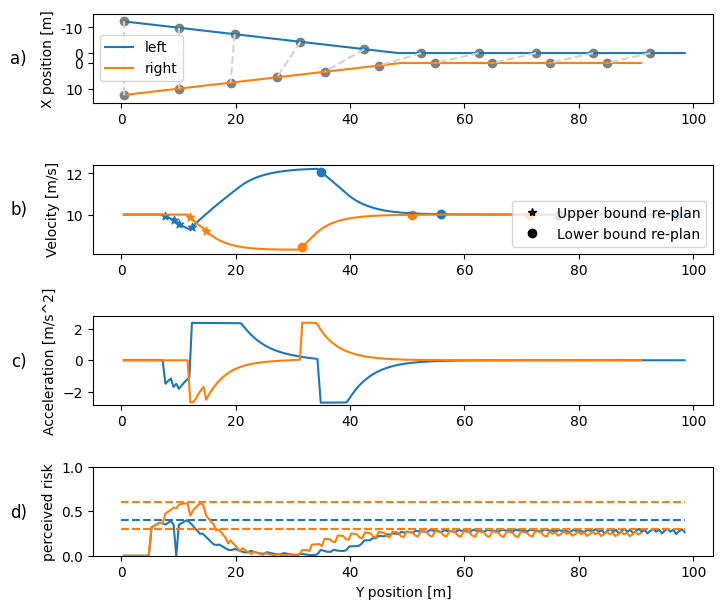}
    \caption{Model behaviour in scenario D. The slight change in $\rho_l$ for the left driver (in comparison to scenario C) resulted in a major change in high-level outcome. Instead of the right driver, the left driver now reaches the merge point first. For details of the notation, see the caption of Figure~\ref{fig:scenarioA}.}
    \label{fig:scenarioD}
\end{figure}

However, the model simulation results show major differences between scenarios C and D (Figures~\ref{fig:scenarioC}~\&~\ref{fig:scenarioD}). As expected, the smaller difference between the left driver's low and high risk threshold resulted in more plan updates. But unexpectedly, this more frequent re-planing resulted in the left driver starting to accelerate and reaching the merge point first. To keep their perceived risk under control, the left driver deviated from their desired velocity to a larger extent than the right driver. This observation can be explained by the fact that high velocities and accelerations do not contribute to risk. The left driver takes whatever action is needed to keep the probability of a collision below their threshold (in this case, high acceleration and high velocity). The slight change in risk thresholds and more frequent re-plans resulted in one of the re-plans initially failing. This triggered a change in the left driver's high-level strategy, they accelerated instead of braked, and this heavily influenced the outcome.

\subsection{Summary Scenarios C and D}
In scenario C, the driver with the higher risk thresholds (the right driver) passed the merge point first. This driver changed their plan at a later moment compared to the other driver. In terms of human behaviour, this can be explained as being more aggressive. The effect of slight changes to the lower threshold was shown to be substantial in scenario D. A small change resulted in a different interaction strategy, making the theoretically more "conservative" left driver arrive at the intersection first. This more conservative driver used high velocities and accelerations to lower their perceived risk even though high velocities would be interpreted by many human drivers as high-risk behaviour. The reason for this seemingly counter-intuitive model behaviour is that the high velocities and accelerations on their own do not contribute to the perceived risk of these modelled drivers.

\FloatBarrier
\subsection{Emergent gap-keeping behaviour for car following}
Although the main focus of our model is on the interactive behaviour of the drivers when approaching the merging point, it also provides insight into their behaviour after the merging conflict is resolved. Specifically, in the four scenarios above, we found that the simulated drivers continued maintaining a gap on the straight section after the merge point. This behaviour was not explicitly programmed and the planner has no cost associated with short time or small distance gaps (a feature frequently used in human driver models~\cite{Naumann2020a, Sadigh2018}). Instead, these distance gaps appear to emerge from the combination of risk perception and a probabilistic belief about the plan of the other driver. 

\begin{figure}[h]
    \centering
    \includegraphics[width=0.9\textwidth]{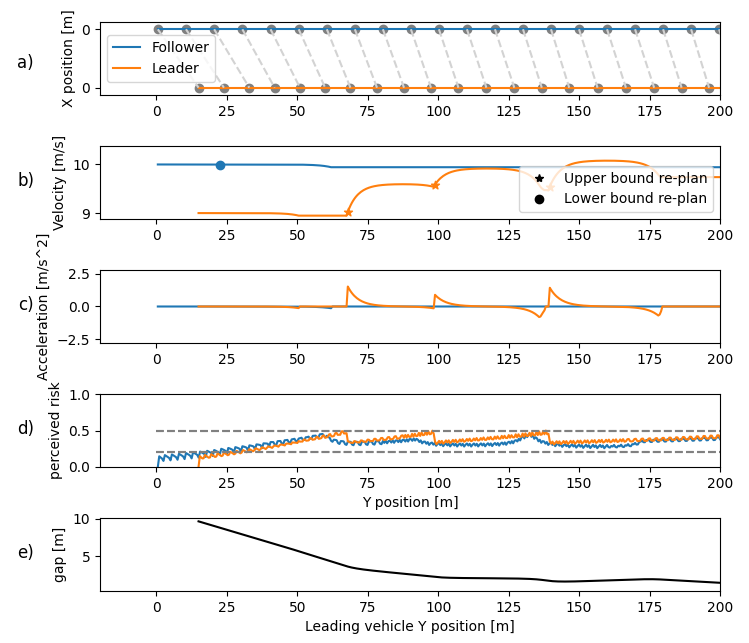}
    \caption{Model behaviour in the straight road scenario. For details of the notation, see the caption of Figure~\ref{fig:scenarioA}. The bottom panel shows the gap between the vehicles as a function of the leading vehicle position. In this scenario, the blue (following) vehicle has a higher preferred velocity than the orange (leading) vehicle. The x-axes have been cropped to the first $200$ meters of the $400$ meter track.}
    \label{fig:gap_emergence}
\end{figure}

To further investigate this effect, we investigated a scenario without a merging point. In this scenario, the drivers drive behind each other on a straight stretch of road ($400~m$). We used the default parameters from Table~\ref{tab:scenario_parameters}, except for the velocity parameters. The leading vehicle has lower desired and initial velocities ($9~m/s$) compared to the following vehicle ($10~m/s$). Figure~\ref{fig:gap_emergence} shows that a steady-state gap emerges after approximately $100$ meters. In this scenario, the leading driver mostly acts to reduce the risk and prevent a collision. 

Although the fact that the leading, not the following, driver mostly acts to maintain this gap is not uncommon for human drivers and has been observed under some conditions~\cite{Xu2021}, it is not the most common behaviour for reducing the risk during car following~\cite{KONDOH2008}. We identified two causes for this model behaviour. First, the belief and risk perception in the model are purely symmetrical. There is no difference in risk between drivers that are in front or behind another, nor is there any difference in believed probability that a driver will accelerate or decelerate. In natural traffic this simplification will not hold, this should be accounted for when extending the model for use in those scenarios. Second, the risk thresholds of both drivers are equal in this example. It can be expected that in other situations, even under the previously mentioned assumption, the driver with the lower risk threshold will act to maintain the gap, as was seen in scenario C. This can be either the leading or the following driver, as was observed in human behavior~\cite{Xu2021, KONDOH2008}.

\begin{figure}[h!]
    \centering
    \includegraphics[width=0.6\textwidth]{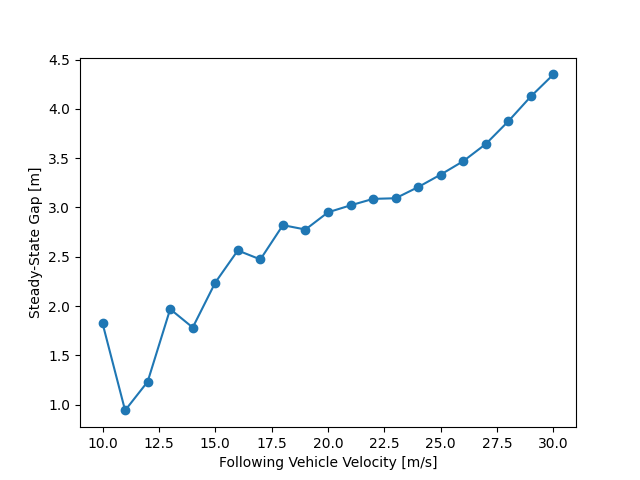}
    \caption{Steady-state gap sizes (averaged over the last second) on a straight road where the following vehicle has a higher preferred velocity. The velocity difference between the vehicle is 10\% and the initial time gap is $1~s$.}
    \label{fig:gap_sizes}
\end{figure}

We investigated the effect of absolute velocities on the resulting steady-state distance gap, where we take the average gap over the final second of simulation as the steady-state gap. We simulated the model behaviour in this scenario for different velocities, every time with a $10~\%$ velocity difference between the drivers, and an initial time gap of $1~s$. We found that the emerging steady-state gap increased linearly with increasing velocities (Figure~\ref{fig:gap_sizes}). This corresponds to human behaviour: the same linear relationship has been previously observed in a study on human gap-keeping behaviour on highways with low speeds~\cite{Piao2003}. 

Our model explains this relationship between velocity and distance gap as follows: The leading driver (orange) is unsure about the future plan of the following driver (blue). It could be possible that the blue driver will accelerate in the near future; In this case, a collision can occur. Because the orange driver keeps its risk below a threshold, it will keep a distance from the blue driver to make sure that its plan does not overlap too much with the possible future positions of the blue driver. Higher velocities, with the same maximum comfortable acceleration, result in a high standard deviation in the belief points. This causes the gap size to increase with velocity.

The mentioned study~\cite{Piao2003} also showed that humans keep larger gaps (approximately $12~m$ to $23~m$ for the same velocity range) compared to our model. We, therefore, conclude that the model qualitatively captures the underlying risk-mitigation mechanism in human car-following behaviour, but needs to be further explored to investigate if fitting the model parameters to human data would also allow it to capture the magnitude of the gap characteristic of human drivers.

\FloatBarrier
\section{Discussion}
\label{sec:discussion} 
In this work, we have proposed a modelling framework for two-way human-human interactions in traffic. We illustrated the utility of the framework by implementing a concrete model based on the framework, targeted at interactive behaviour in a simplified merging situation. We investigated the model's behaviour in four scenarios, one where the drivers are not on a collision course, one where they are, and two where we investigated the effects of the model parameters. The model captures the actions of two drivers who 1) successfully resolve merging conflicts without collisions, 2) increase safety margins that are clearly too small (a $20~cm$ gap) for human drivers, and 3) exhibit individual conservative and aggressive behaviour, based on physically meaningful model parameters: their risk thresholds. In all scenarios, the model behaves in a plausible way that corresponds to intuitions about human interactive behaviour in merging conflicts.

Furthermore, from the model's underlying principle (the notion of risk combined with the probabilistic belief about the other driver's plan) plausible behaviour emerged outside of the situations we developed and tuned the model for. Specifically, a realistic gap-keeping behaviour emerged, where the drivers keep larger distance gaps at higher velocities, as humans do~\cite{Piao2003}. This behaviour was observed even though no distance or time gap-related costs are incorporated in the model. These results show that the proposed model framework is a promising novel approach for modelling two-way multi-agent interactions in traffic.

Modelling interactions in traffic has both practical and fundamental applications. In practice, a modelling framework like the one we propose could aid the development of autonomous vehicle controllers that aim to increase acceptability and safety in interactive scenarios. More fundamentally, such modelling, even when limited to an isolated traffic scenario, could contribute to gaining fundamental knowledge of human behaviour by highlighting the cognitive mechanisms humans use when interacting with each other. Our novel framework addresses the limitations of existing modelling and control approaches, among which game-theoretic models and interaction-aware controllers, because it explicitly incorporates communication and two-way interaction. Furthermore, our model framework does not make strong assumptions about human behaviour, such as the assumption that humans are rational utility maximizers. We hope that the initial exploration of the model framework presented here can spark a new strain of interaction modelling research.

\subsubsection*{Similar Approaches}

Among existing approaches to modelling traffic interactions, by far the most explored one is game theory. For example, for an extensive review of game-theory-based lane-changing models, see~\cite{Ji2020a}. What is similar to our framework, is that game theory aims at modelling two-way interactions instead of modelling only one driver responding to another (for examples, see~\cite{Meng2016,Tian2019,Zhang2018,Coskun2019,Liu2007,Kita1999}). What is different, is that our approach is not limited by two main assumptions (rationality, and lack of communication), and --for the majority of GT approaches-- a focus on decision-making without describing operational behaviour. Finally, and more conceptually, game-theoretic models implicitly approach traffic interactions as a competition, while in our framework the agents have a joint primary objective (interaction safety) that makes the interaction a joint, cooperative effort.

In contrast with game theory, our approach explicitly incorporates communication between drivers. Although there are similarities with game theory, for example, our case study uses the same modality of communication as many game theoretical approaches, position and velocity observations (e.g.,~\cite{Liu2007, Li2018}, for an overview, see~\cite{Ji2020a}). There are two fundamental distinctions in how we approach communication with respect to game theory.

First, the communication in our framework allows drivers to construct and update a belief about the other vehicle's plan without the need for any prior information about the other driver. This is a fundamental contradiction with game theory where players are assumed to know each other’s utility functions (at least partially) beforehand. Therefore, in game theory, communication is not necessary because players can reason about what the other player is going to do to maximize their utility given the current state. The observations of position and velocity are only used to determine the state of the world. While in our model, position and velocity are used to convey information about the intention of other drivers.

Second, in game theory, observations are not “remembered”. They only serve to determine the \underline{current} state, which is enough to reason about the other players’ actions. Previous states are irrelevant. This is also known as the Markov condition or assumption. While in our work, the history of communication is kept in the belief about the other driver’s intentions. Thus, the belief about a driver's future actions is based on its recent behaviour, not only on the current state. Some approaches combine game theory with an online estimation of the other player’s utility function, thereby indirectly basing the belief about future actions (which directly depends on the utility function) on recent behaviour (e.g.,~\cite{Schwarting2019, Sadigh2018}). However, in these approaches, the conveyed information is not regarded as intentional communication. Furthermore, these approaches only estimate part (e.g., a single parameter) of the utility function online, the rest is assumed to be known a priori.

Another modelling concept that bears resemblance to our approach is that of \textit{Belief-Desire-Intent (BDI) modelling}. BDI modelling is based on the philosophical work of Bratman~\cite{Bratman1987} and models single agents that have a belief, a desire (goal), and an intent (plan). Many implementations of BDI models have been proposed for different applications~\cite{DeSilva2020}. The BDI framework and our CEI framework share the concepts that agents construct a (probabilistic) belief about other agents and the world, and then make a plan based on that belief to reach a final goal. The BDI framework, however, was not indented to account for interactions. It is primarily a model framework for individual agents that perform individual tasks. It therefore also does not incorporate communication but instead updates its beliefs based on changes that occurred in the world.

Finally, an important concept that can be complementary to the CEI-model framework, and bears resemblance to the BDI framework is the concept of \textit{Theory of Mind (ToM)}~\cite{Premack1978} (for examples of applications to human-robot interaction, see~\cite{Scassellati2002, Devin2016}). ToM is a psychological concept that assumes humans have an internal model of the beliefs, goals, and intentions of other humans in an interaction. Thereby, having the ability to reason about want other humans want, and how they will try to achieve that goal. This idea that humans understand the mechanisms behind the actions and beliefs of others could be used in an implementation of our proposed CEI-model framework, which, in principle, only requires humans to form a basic belief about the future movements of others. As an example, the implementation of the CEI-model in the case study assumes drivers predict where the other driver is going, not why they are doing that. A complete ToM model could extend this belief about future actions of the other, with beliefs about their beliefs and goals. Implementing a CEI-based model with an internal ToM model is an interesting avenue for future research.

Besides these different types of modelling approaches, recently a great deal of effort was put into approaches for controlling (autonomous) vehicles in merging scenarios (e.g.~\cite{Sadigh2018,Schwarting2019,Nishi2019}). Although the underlying techniques (such as finding a policy by optimizing some utility function) can be similar, the goal of these approaches is very different. While modelling approaches (such as ours) aim to best describe human behaviour. Control approaches aim to find a safe and optimal solution to a control problem. Game theory can therefore be very suitable for use in control approaches (as was done in~\cite{Sadigh2018,Schwarting2019,Yu2018a}).

Two recent works on modelling come close in scope to this work. In 2022, Markkula et al. proposed a modelling approach for individual agents in a driver-pedestrian interaction rather than multiple agents in a driver-driver interaction~\cite{Markkula2022}. Using different versions of a model that incorporates a variety of concepts from psychology, with varying levels of complexity, they conclude that "modelling of human road user interaction is a formidable challenge". Similar to our work, their findings suggest that the problem cannot be solved with simple rational models. Besides that, accounting for specific, previously unexplained, phenomena observed in human interactive behaviour could only be done using complex cognitive models. These conclusions resonate with our argument that the development of new model frameworks that go beyond game theory and the assumption of one-way interaction is a necessary step to improve our understanding of human traffic interactions.

Secondly, in 2014, Wan et al. also proposed an approach to model vehicle-vehicle interactions on merging ramps~\cite{wan2014}. As in our work, they specifically address the influence vehicles have on each other. Their (and our) work, therefore, differs from traditional driver models that usually describe a single driver responding to -- but not influencing -- other traffic. Another similarity between our proposed framework and the work by Wan et al. is that we both explicitly consider communication between vehicles. However, the model proposed by Wan et al. specifically targets congested traffic and uses different mathematical models for vehicles that have different roles in the interaction (i.e., they determine who will lead, follow, and merge a priori). Wan et al. also do not consider individual differences between drivers.

\subsubsection*{Framework Extensions}
Although we have only demonstrated our proposed model framework for a simple merging scenario with two vehicles, it could easily be extended to more vehicles or traffic interactions with other types of participants. The underlying reason is that while we put the model's bounding box around the complete interaction, the drivers within the model are strictly separated; the only component connecting the two drivers is communication (Figure~\ref{fig:model}). This has two main advantages. First, communication in our framework is based on observable signals (e.g., turn indicators or velocity). This means that sending and receiving communication can easily be shared between multiple drivers, i.e., the communication is broadcast to all surrounding road users rather than sent directly to one of them. For that reason, the model framework can be extended to any number of drivers without requiring a redesign. Second, because the drivers are separated, it is possible to swap one of the drivers in the model with another type of agent, for example, a pedestrian. This would require adding the agent type to the observed communication, but since this is also an observable feature, it would not make the model more complex. 

One could even go as far as replacing one of the agents in the model with a non-model agent altogether. This could, for example, be used to let a real human interact with the model in a driving simulator (this would require an optimized model implementation capable of running in real-time). This in turn would allow for the possibility of human drivers subjectively evaluating the ability of the model to describe natural interactions. Alternatively, a model could be used to evaluate autonomous vehicle controllers by letting the model interact with such a controller. Another potential extension useful for AV development is integrating the model into an AV controller to help it make decisions with an online evaluation of potential outcomes of an interaction.

We believe our model could also be adapted to other types of human-human interaction tasks. An example of such a task is cooperative bottle reaching, for which a communication model was developed in~\cite{Pezzulo2013}. The task in~\cite{Pezzulo2013} is similar to our task in that it constitutes a joint effort for which communication and action take place along the same channel (velocity/acceleration in our case). The main difference between our model framework and the communication model in~\cite{Pezzulo2013} is that we target the interaction dynamics, in which we assume communication plays an important role, instead of targeting to model the communication as a stand-alone feature. 

\subsubsection*{Limitations and Future Work}
Both the specific model implementation and the general modelling framework have important limitations. To start with the former, the model used for the simplified merging scenario uses very simplistic implementations for all components. The plan is based on desired velocity and acceleration alone. The beliefs are one-dimensional and assumed to be Gaussian distributions. The communication is assumed to be perfect (continuous without any noise), and only based on implicit cues. And finally, the risk is only based on collision avoidance, not influenced by high velocities or accelerations. In future implementations of the model, these limitations need to be addressed. However, it is important to first identify which of these limitations (if any) play a role in the model's ability to accurately reproduce human-human interactions. This could be done by comparing the model to data on human-human interactions gathered in a driving simulator experiment. 

Another limitation of the current model implementation lies in the updates of the belief function. The assumption that the likelihood function (used for the Bayesian updates) has a known and fixed standard deviation results in the fact that every update reduces the standard deviation of the posterior, even if the new information contradicts the current belief. This is counter-intuitive: contradicting information (incoming through communication) should increase the variability of the belief, not decrease it. Put differently, if another person or driver sends unclear communication about what they are going to do by alternating between accelerating and braking, one should keep all options open, not decrease the standard deviation of the predicted position after a couple of seconds while shifting the mean around on every time step. How to properly address this limitation remains an open question.

Finally, the model's satisficing-based decision-making can result in unstable outcomes for high-conflict scenarios. When re-planning, the drivers in the model will first search for a new solution close to the previous solution. For example, if the previous plan was to brake, the driver will first explore if braking harder will satisfy the new constraint. Only if this optimization fails, the driver will explore other strategies (i.e., acceleration) to lower the perceived risk. This drastic change in high-level behaviour is thus triggered by the first optimization failing. Therefore, slight numerical or temporal differences in this optimization can lead to different high-level outcomes, especially for situations that are highly symmetrical (e.g., when drivers have very similar parameters and none of the vehicles has a clear kinematic advantage). This was already observed in scenario D, where a slight change in model parameters caused a different outcome, but a similar outcome change could also result from changes in the type of numerical optimization solver or its parameters. One way of addressing this sensitivity is to make the model stochastic: introducing variability in the model's behaviour will make the outcome in high-conflict scenarios inherently stochastic and therefore could help to make it less sensitive to small external perturbations.

Adding stochasticity also addresses the main limitation of the overall framework, which is that currently, the framework is fully deterministic: with the exact same parameters (for model and solver), the model will always produce the same behaviour. This is inconsistent with the substantial behavioural variability that humans exhibit in traffic~\cite{Kurtc2020}. We see multiple possible ways of introducing stochasticity in the framework to account for this. To name two: adding stochasticity could be done in the receiving of communication (translating perceptual information to an updated belief) by using evidence accumulation mechanisms~\cite{Zgonnikov2020} or additive noise, or by including noise directly in the risk perception. However, more work is needed to determine the best approach.

A second limitation of the overall framework concerns improvements and redesigns of the model. Although the different components in the framework are separated, which should allow for easy redesign of parts of the model, they do depend on each other. This could mean that when redesigning one aspect of the model, a redesign of another aspect is inevitable. As an example, in the case study, we used velocity and position as the means of communication. These values are directly used in the belief update. However, if we would change the communication component of the model, the belief and its update also need to be changed. This is an important consideration when starting a redesign of the model since this could be the case for more components.  

Finally, event-based triggering of the re-plan based on perceived risk results in an uneven computational requirement from the model: some time steps may take significantly more time to compute than others. A result of this is that our current implementation of the model cannot run in real-time. Instead, we used offline simulation for the case study. This could pose a problem when an experiment needs to be performed where the model interacts directly with a human.

Although the presented case study shows promising results, there is much future work to be done on the proposed framework. In addition to accounting for stochasticity in human behaviour and optimizing the runtime performance of the model, a necessary next step is to compare the model to human-human interactive behaviour. However, even validating single-driver models that do not incorporate interactions is already a complex task~\cite{Siebinga2022d}, therefore comparing our model to human-human interaction data requires a separate detailed investigation.

\section{Conclusion}
In this paper, we proposed a novel modelling framework to model human-human driving interactions. The key insight underlying this framework is the focus on the joint behaviour of the drivers during the interaction, rather than the isolated behaviour of a single driver. The framework explicitly includes communication between drivers and mutual influences (two-way interaction). We implemented the model for a simplified merging scenario and investigated its behaviour in four scenarios. We conclude the following:

\begin{itemize}
    \item The model avoids impending collisions via plausible driver-driver interactive behaviours;
    \item Changing the risk threshold parameters per driver results in changes in behaviour that can be interpreted as more aggressive or conservative;
    \item Velocity-depended gap-keeping behaviour emerges from the combination of risk-based planning and a probabilistic belief about other drivers' plans. With this behaviour, the model shows a fundamental aspect of human driving behaviour, without it being explicitly programmed; 
    \item The proposed model framework is a promising novel approach for modelling two-way multi-agent interactions in traffic.
\end{itemize}

\section*{Acknowledgement}
The authors thank Nissan Motor Co. Ltd. for funding this work.

\newpage
\bibliographystyle{ieeetr}
\bibliography{main}

\end{document}